 \documentclass[prl,aps,twocolumn,showpacs]{revtex4}
                                                               \usepackage[dvips]{graphicx}
\usepackage{dcolumn}
                                                                           
\newcommand{\be}{\begin{equation}}
\newcommand{\ee}{\end{equation}}
\newcommand{\bea}{\begin{eqnarray}}
\newcommand{\eea}{\end{eqnarray}}
\newcommand{\nn}{\nonumber}

\begin{document}
\topmargin=-20mm

\title{On the Electronic Transport Mechanism  in  Conducting Polymer Nanofibers} 

\author{Natalya A. Zimbovskaya$^\dagger$,  Alan T. Johnson, Jr.$^\ddagger$, and Nicholas J. Pinto$^\dagger$}

\affiliation{$^\dagger$Department of Physics and Electronics, University of Puerto Rico at Humacao, Humacao, PR 00791, and $^\ddagger$Department of Physics and Astronomy, University of Pennsylvania, Philadelphia, PA   19104-6396 }

\date{\today}

 \begin{abstract}
Here,  we present theoretical analysis of electron transport in polyaniline based (PANi) nanofibers assuming the metalic state of the material. To build up this theory we treat conducting polymers  as a special kind of granular metals, and we apply the quantum theory of conduction in mesoscopic systems to describe the transport between metallic-like granules. Our results show that the concept of resonance electron tunneling as the predominating mechanism providing charge transport between the grains is supported with recent experiments on the electrical characterization of single PANi nanofibers. By contacting the proposed theory with the experimental data we estimate some important parameters characterizing the electron transport in these materials. Also, we discuss the origin of rectifying features observed in current-voltage characteristics of fibers with varying cross-sectional areas.
  \end{abstract}

\maketitle

Starting from their discovery, conducting polymers  undergo intense research. These materials are significant mostly due to various possible applications in fabrication of nanodevices, for polymer based devices should have advantages of low cost and flexible, controlled chemistry \cite{one}. Also, there are some unsolved problems concerning physical nature of charge transfer mechanisms in conducting polymers making them interesting subjects to fundamental research.
Recently, we reported of fabrication of doped 
polyaniline/polyethylene oxides (PANi/PEO) nanofibers using electrospinning technique \cite{two,three}. 
The details of the electrospinning setup used for the  fibers fabrication are described elsewhere \cite{two,four}.

Single fiber electrical characterization was carried out at room temperatures $(T \sim 300$ K). 
It was revealed in the experiments that current-voltage curves for conducting PANi fibers could be non-ohmic. Field-effect transistor (FET) saturation behavior was observed in the device made out of two nanofibers with the diameters 300 nm and 120 nm, respectively, placed across two gold electrodes \cite{two} on a doped Si/SiO$_2$ wafer in a FET configuration. As well, nonlinear current-voltage curves were obtained in trasport experiments  on thinner nanofibers \cite{three}. Some characteristics  appeared to be asymmetric with respect to the reversal of the applied bias voltage polarity, especially those for fibers of varying diameter. We do believe that these experimental results contain important information concerning physical mechanisms of the charge transport in conducting polymers. In the present work we concentrate on the theoretical analysis of the previous experimental results to recover this information.


 Conducting chemically doped polymers are known to be very inhomogeneous.  In some regions polymer chains are disordered,  forming amorphous poorly conducting substance. In other regions the polymer chains are ordered and densely packed \cite{five,six}. These regions behave as metallic-like grains embedded in a disordered amorphous  environment. This gives grounds to apply the model of a granular metal \cite{seven} to describe such materials.  The fraction of metallic-like islands in bulk polymers varies depending on the details of synthesis process. Presumably, in actual samples the metallic grains always remain separated by poorly conducting disordered regions and do not have direct contacts.
 Within this model electronic states are assumed to be delocalized over the grains, therefore electrons behave as conduction electrons in conventional metals. Their motion inside the grains is diffusive with the diffusion coefficient $ D =  \frac{1}{3} v_F^2 \tau \ ( v_F $ is the Fermi velocity, and $ \tau $ is the scattering time).

Recently, Prigodin and Epstein (PE)  suggested that the grain-to-grain transport mostly occurs due to the resonance tunneling of electrons. The latter is provided with intermediate states on the  chains connecting the grains  \cite{eight}. In the present work we further develop the PE approach to the charge transport in conducting polymers, and we show that it brings results consistent with those obtained in  transport experiments on individual PANi nanofibers \cite{two,three}.

Within this approach we immediately see  a similarity in electron transport mechanisms in  conductive polymers and 
those in molecular wires connecting metal leads \cite{nine}. 
In both cases the  transport is mostly provided by intervening "bridges" giving rise to intermediate states for the electron tunneling. Employing the granular metal model we can treat metallic islands as "leads" connnected with single-site bridges. This simple structure of the bridges is proposed in PE theory. We also explain the issue below.  A systematic theory of conductance in quantum molecular wires is developed during the last two decades.  In further analysis  we employ this theory. Accordingly, we write  an electric tunneling current flowing between two crystalline domains $I_{ig} $ in the form \cite{ten}:
    \be  
  I_{ig} = \frac{2e}{h} \int_{-\infty}^\infty dE T_{eff}(E) \Big[ f_1(E) - f_2(E) \Big ].
  \ee
 Here, $ f_{1,2} (E) $ are Fermi functions taken with different contact chemical potentials for the grains: $ \mu_1 $ and $ \mu_2, $ respectively. The chemical potentials differ due to the voltage $ \delta W $ applied between the grains \cite{eleven}: 
    \be 
  \mu_1 = E_F + (1 - \eta) e \delta W; \qquad
 \mu_2 = E_F - \eta e \delta  W.  
  \ee
 The dimensionless parameter $ \eta $ characterizes how the voltage is divided between the grains,  $E_F$ is the equilibrium Fermi energy of the system, and $T_{eff} (E)$ is the electron transmission function.


In general, the electron transmission function $ T_{eff} (E) $ includes a term $ T(E) $ describing coherent transport by electron tunneling along with inelastic contribution originating from  motions in the medium between the grains. These motions, especially nuclear motions in the resonance chain, could cause the electronic phase breaking effect. In our case "the bridge" is a single electron level situated between the grains.

For the considered problem the electron transport is a combination of tunneling through two barriers (the first one separates the left metallic domain from the intermediate state in the middle of the resonance chain, and the second separates this state from the right grain, supposing the transport from the left to the right) affected with the inelastic scattering at the bridge, as shown below. The barriers are represented by the squares, and the triangle in between imitates a scatterer coupling the bridge to a dissipative electron reservoir. 

{\small
\vspace{14mm}

\begin{picture}(0,0)(-25,0) 
\put(10,0){\framebox(30,30)}
\put(140,0){\framebox(30,30)}

\put(60,4){\line(1,0){60}} 
\put(60,4){\line(5,4){30}}
\put(120,4){\line(-5,4){30}}
\put(70,4){\line(0,-1){30}}
\put(110,4){\line(0,-1){30}}
\put(66,-4){\vector(0,-1){15}} \put(54,-14){$a_4'$}
\put(74,-19){\vector(0,1){15}} \put(77,-14){$a_4 $}
\put(106,-19){\vector(0,1){15}} \put(94,-14){$a_3$}
\put(114,-4){\vector(0,-1){15}} \put(117,-14){$a_3'$}

\put(10,15){\line(-1,0){25}}  
\put(40,15){\line(1,0){34}}
\put(140,15){\line(-1,0){34}}
\put(170,15){\line(1,0){25}}
\put(60,-26){$ ....................... $}
\put(72,-44){Reservoir}
\put(67,-35){\normalsize 4} \put(108,-35){\normalsize 3}
\put(-21,12){\normalsize 1} \put(197,12){\normalsize 2}

\put(-10,20){\vector(1,0){15}} \put(-5,25){$b_1$}
\put(5,10){\vector(-1,0){15}} \put(-5,0){$b_1'$}
\put(45,20){\vector(1,0){15}}  \put(46,25){$a_1$}
\put(60,10){\vector(-1,0){15}} \put(46,0){$a_1'$}
\put(135,20){\vector(-1,0){15}} \put(125,0){$a_2'$}
\put(120,10){\vector(1,0){15}}  \put(125,25){$a_2$}
\put(190,20){\vector(-1,0){15}} \put(180,25){$b_2$}
\put(175,10){\vector(1,0){15}}  \put(180,0){$b_2'$}

\end{picture}}
\vspace{18mm}

An electron could be injected into this system, and/or leave from there via four channels indicated  in this schematic drawing. Incoming particle fluxes $ (J_i) $ are related to those outgoing from the system $ (J_j') $ by means of the transmission matrix  T \cite{twelve,thirteen}:
    \be 
 J_j'= \sum_i T_{ji} J_i, \qquad 1 \leq i, \ j \leq 4.
   \ee
  Off diagonal matrix elements $ T_{ji} (E) $ are probabilities for the electron to be transmitted from the channel $ i $ to the channel $ j, $ whereas diagonal matrix elements $ T_{ii} (E) $ are probabilities for its reflection back to the channel $ i $ \cite{fourteen}. To provide the charge conservation, the net particle flux in the channels connecting the system with the reservoir, must be zero. So, we have:
   \be 
  J_3 + J_4 - J_3'- J_4' = 0. 
   \ee
   We have no grounds to believe that there is difference between the channels 3 and 4, therefore we assume $ J_3 = J_4. $ Also,  in the following calculations we put $ J_2 = 0, $ presuming no income flux in the channel 2.

Now, we employ (3) to express outgoing fluxes in terms of incoming ones, and we substitute the results into the equation (4). This gives:
    \be 
  J_3 = J_4 = J_1 \frac{K_1(E)}{2 - R(E)} .
  \ee  
  Here,
    \bea
 && K_1 (E)  =  T_{31} + T_{41}; \nn \\ \nn  \\
&& R(E) = T_{33} + T_{44} + T_{43} + T_{34}.  
   \eea
  The transmission function $ T_{eff} (E) $ relates the particle flux outgoing from the channel 2 to that incoming to the channel 1, namely:
  \be 
  J_2'= T_{eff} (E) J_1.
  \ee
  So, using Eqs.(3),(5) we obtain:
     \be 
  T_{eff} = T(E) + \frac{K_1(E) K_2(E)}{1 -R(E)},
   \ee
  where
  \be 
  T(E) = T_{21}; \qquad
 K_2 (E) + T_{23} + T_{24}.
     \ee
   The functions $  T(E),\ K_{1,2} (E) $ and $ R(E) $ are expressed in terms of the matrix elements of the scattering matrix  $ S $ relating outgoing wave amplitudes  $b_1',b_2',a_3',a_4'$ to the incident ones $b_1,b_2,a_3,a_4: \ T_{ij} = |S_{ij}|^2. $ In the considered case the $ S $ matrix takes the form \cite{fourteen,fifteen}:
    \be 
S = Z^{-1}
\left (
\begin{array}{cccc}
 r_1 + \alpha^2 r_2 & \alpha t_1 t_2 & \beta t_1 & \alpha \beta t_1 r_2
  \nn \\
  \alpha t_1 t_2 & r_2 + \alpha^2 r_1 & \alpha \beta r_1 t_2 &  \beta t_2 
\nn \\
\beta t_1 & \alpha \beta r_1 t_2 & \beta^2 r_1 & 
 \alpha r_1 r_2 - \alpha
\nn \\
\alpha \beta t_1 r_2 & \beta t_2  &  \alpha r_1 r_2 -  \alpha  & \beta^2 r_2
  \nn 
\end{array}
\right )
   \vspace{1mm}
                 \ee
  where $ Z = 1 - \alpha^2 r_1 r_2 ; \ \alpha = \sqrt {1 - \epsilon} ; \ \beta = \sqrt{\epsilon}; \ r_{1,2} $
 and $ t_{1,2} $ are the amplitude transmission and reflection coefficients for the barriers  $ ( |t_{1,2}|^2 + |r_{1,2}|^2 = 1); $ and the parameter $ \epsilon $ characterizes the dephasing strength. This parameter takes values within the range $ [0,1], \ \epsilon = 0 $ corresponding to the completely coherent, and $ \epsilon  = 1 $ to the fully incoherent transport, respectively.

When the bridge is detached from the dephasing reservoir $ T_{eff} (E) = T(E).$ In this case we can employ a simple analytical expression for the electron transmission function  \cite{sixteen,seventeen,eighteen}:
    \be  
  T(E) = 4 \Delta_1 (E) \Delta_2 (E) |G (E)|^2,
  \ee
 where:
   \be 
 \Delta_{1,2} (E) \equiv - \mbox {Im} \Sigma_{1,2} (E) = \frac{V_{1,2}^2}{E - \epsilon_{1,2} - \sigma_{1,2}}.
  \ee
  Self-energy terms $\Sigma_{1,2}$ appear due to the coupling of the metallic grains ("leads") to the  bridge, and $V_{1,2} $ are the corresponding coupling strengths. The form of these terms originates from a frequently used  model where the grains are simulated with semiinfinite homogeneous chains. The parameters $ \sigma_{1,2} $ represent  self-energy corrections of the chains. In further analysis we estimate $ \Sigma_{1,2} $  at the energies close to the Fermi energy $ E_F, $ which brings the result: $ \Sigma_{1,2} \approx - i V_{1,2}^2/\gamma_{1,2} $ \cite{thirteen}. Here, $ \gamma_{1,2} $ is the coupling strength in the chains imitating the metallic grains.

The retarded Green's function for a single site bridge could be approximated as follows \cite{nineteen}:
    \be 
  G(E) = \frac{1}{E - E_1 + i \Gamma}
  \ee
 where  $ E_1 $ is the site energy. The width of the resonance level between the grains is  described with the parameter $ \Gamma = \Delta_1 + \Delta_2. $

Comparing the expressions (8) and (11) in the limit $ \epsilon = 0 $ we arrive at the following expressions for the tunneling parameters $ \delta_{1,2} (E): $
  \be 
 \delta_{1,2} (E) \equiv t_{1,2}^2 (E) = \frac{2 \Delta_{1,2}}{\sqrt{(E - E_1)^2 + \Gamma^2} }.
  \ee
 
Using this result we easily derive the general expression for the electron transmission function: 
  \be 
  T_{eff}(E) = \frac{g(E) (1 + \alpha^2)[g(E)(1+\alpha^2) + 1 - \alpha^2]}{[g(E)(1 - \alpha^2) + 1 + \alpha^2]^2}
  \ee
 where $g(E) = t_1 (E) t_2 (E). $

In further calculations we keep in mind that
the transmission coefficient for the resonance tunneling is determined with the probability of finding the resonance state.  The latter is estimated as $ T \sim \exp (-L/\xi ) $ where $ L $ is the average distance between the interacting grains, and $ \xi $ is the localization length for electrons \cite{eight}.  It takes values small compared to unity but much greater than the transmission probability for sequental hoppings along  the chains $ T_h \sim \exp (-2L/\xi ) $ \cite{tventy}. The probability for existence of a resonance state at a certain chain is rather low, so only a few out of the whole set of the chains  connecting two grains are participating in the process of  intergrain electron transport.  Therefore one could assume that any two metallic domains are connected with a single chain providing an intermediate state for the resonance tunneling. All remaining chains could be neglected for they poorly contribute to the transport  compared to the resonance chain.

Realistic PANi nanofibers prepared via electrospinning have diameters in the range $ \sim 20 \div 100 nm$  and lengths of the order of several microns. This is significantly greater than typical sizes of metal-like grains and intergrain distances, which take on values $ \sim 5 nm $ \cite{tventyone}.  So, we  treat a nanofiber as a set of parallel channels for the tunneling current, any channel being a sequence of metallic domains connected with resonance chains. The total current in the fiber is the sum of the contributions from all these channels. We assume  these contributions to be equal. Also, the voltage $ W $ applied across the whole  fiber is a sum of contributions $ \delta W $ from sequental pairs of grains along a single channel. 
 We assume as the fist approximation $ \delta W \sim WL/L_f $ where $ L_f $ is the fiber length. Since we presume the current to be conserved in any single channel, we can identify the intergrain current $I_{ig} $ given by (1) with the current in the channel. Using the above approximation for $ \delta W $ we express the current in the channel in terms of the net voltage $ W $ instead of $ \delta W. $

For certain doping and crystallinity rates the number of channels is proportional to the fiber cross-sectional area. In reality, dedoping of the fibers due to the contact of their surfaces with atmospheric gases  affects the number of the working channels $n$. This occurs because the electron localization length $ \xi $ decreases in the dedoped surface layers, bringing a crucial fall in the conduction through the channels which run there. As a result these channels become insulating. The relative number of the insulating channels is greater in thinner nanofibers, therefore their conduction (the latter is proportional to the number of working channels) is much stronger reduced due to the dedoping processes, than the conduction of thicker samples. 

Conducting samples studied in the experiments of \cite{three} included $ 70 nm $ diameter nanofiber (Sample 1), and a pair of $ 18 nm$ and $ 25 nm $ diameter fibers connected in parallel (Sample 2). PANi nanofibers whose diameters were smaller than $ 15 nm $ appeared to be insulating due to dedoping, as reported in \cite{three}. Therefore we assume that outer layers whose thickness is $ \sim 8 nm $ are insulating in all fibers used in the experiments.  Accepting the value 5 nm to estimate both average grain size and intergrain distance and keeping in mind the effect of dedoping, we find that the 70 nm fiber in the experiments of \cite{three} could include about $ 30 \div 40 $ conducting channels and we could not expect more than 2 working channels in the pair of fibers $ S2. $
This gives the ratio  of conductions $ \sim 15-20  $ which is greater than the ratio of cross-sectional areas of the samples  $( A_1 / A_2 \approx 5).$ The difference originates from the stronger dedoping of thinner fibers. 

To calculate the current in the whole nanofiber $ I $ we multiply the channel in a single current by the number of working channels $ I = n I_{ig}. $
 In calculations  we consider $|V_{1,2}| << |\gamma_{1,2}|$ because the coupling strength of the grains to the intermediate site in between is proportional to a small factor $ \exp (-L/\xi).$  As for the magnitude of the coupling strengths $\gamma_{1,2} ,$ we can roughly estimate them as $ 1 eV. $ This value is typical (in order) for covalent bonds. To provide a good match of our theory with the experimental data we employ the least square procedure. This brings $ V_{1,2} = 3.0 meV $ for the $ 70 nm $ fiber (Sample 1) for $ n = 35, $ and $ V_{1,2} = 2.4 meV \  (n = 2) $  for the pair of thinner fibers (Sample 2).
  
The value of the dephasing parameter $ \epsilon $ is determined with the interplay between the characteristics of electron-phonon interaction and those of the electron coupling to the leads. When the latter is weak, as in the considered case, $ \epsilon $ should depend on energy $ E $  exhibiting peaks which correspond to vibrational modes in the resonance chain. This was shown in recent papers \cite{eighteen,nineteen} basing on the nonequilibrium Green's function formalism.  These peaks should be revealed in the transmission function $ T_{eff} (E) $ bringing step-like features in the current-voltage characteristics. The absence of such features in the experimental curves reported in the work \cite{three} gives grounds to believe that in these experiments the relevant energy range does not include phonon harmonics associated with the vibrational modes. Therefore, in our calculations we treat $ \epsilon $ as a constant. At a  distance from the peaks, the value of $ \epsilon $ is mostly determined with the strength of stochastic motions in the resonance chain, so this value depends on temperature. Basing on the results of some earlier works (see e.g. \cite{thirteen}) concerning electron transport through molecular bridges at room temperature, we assume $ \epsilon = 0.05 $ in our calculations.

\begin{figure}[t]
\begin{center}
\includegraphics[width=8.8cm,height=4.4cm]{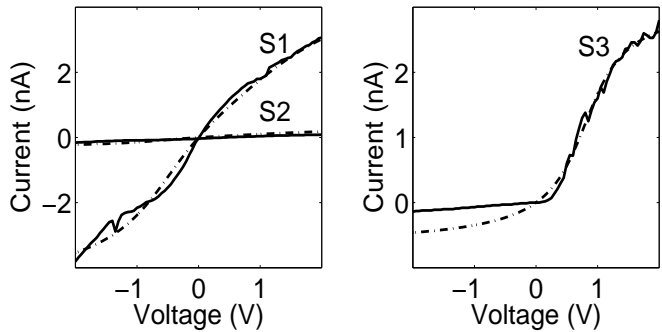}
\caption{
Calculated  (dash-dot lines) and experimental (solid lines) current (nA) -- voltage (V) characteristics for PANi/PEO nanofibers (T = 300 K). Experimental characteristics were reported in \cite{three}.}   
\label{rateI}
\end{center}
\end{figure}

Our results are shown in the Fig.1. The curves S1, S2
present current-voltage characteristics for thick $70 nm $ nanofiber (S1), and the pair of thin fibers connected in parallel (S2). In plotting the strongly asymmetric curve $S3 $ we put $ n = 1, $ and we obtained $ V_{1,2} = 3.2 meV. $ The division parameter value is taken $ 0.35 $ for the curve S1.  This value is chosen to reproduce slight asymmetry of the corresponding experimental curve. On the contrary, to plot the curve S3 we assume $ \eta = 0.02 $ which represents very asymmetric voltage division along the resonance chain.

 The asymmetric curve S3 looks like experimental volt-ampere characteristic reported in \cite{three} for the fiber whose diameter was sharply changing from  70 nm to 20 nm near the middle. The latter did show asymmetric strongly rectifying volt-ampere curves. We conjecture that sudden change in the fiber diameter produces significant changes in the voltage distribution along that segment of the fiber where its diameter changes, affecting the voltage division between the adjacent metallic grains located there,  so that the major part of the voltage fits the thinner segment. Assuming for simplicity that the change in the fiber diameter happens along the segment whose length has the same order as intergrain distances, we can roughly estimate the voltage division comparing cross-sectional areas of the segments. For the fibers used in the experiments of \cite{three} this ratio is $ 12.5. $ Recalling the effects of dedoping on both segments we estimate that voltage division between the segments could be $ \sim 25 : 1 $ which gives for the division parameter at the interface separating the segments estimation: $ \eta \sim 0.04. $ This is reasonably close to the value $ \eta = 0.02 $ used in plotting the curve S3.

 Presented theoretical volt-ampere characteristics reasonably match the curves obtained in the experiments. This justifies the accepted concept of the electron transport in the polymeric granular metals. Also, basing on our results we can estimate values of the coupling strengths  $ V_{1,2} .$ 
 We conclude that the coupling strength  of metallic domains to the intermediate state in PANi fibers could be estimated as $ 2.0 \div 3.5 \,meV. $

Finally, in this paper we apply theory of electron transport through molecular wires to analyze transport characteristics obtained in experiments on PANi nanofibers. Demonstrated agreement between the presented theoretical results and experimental evidence  proves that electron tunneling between metallic-like grains through intermediate states at the resonance chains really plays a significant part in the transport in the conducting polymers provided that they are in the metallic state. This is a novel result which helps to achieve better understanding of physics of metallic state of conducting polymers. Further and more thorough analysis based on the concepts used here promises to bring more  information concerning physical mechanisms and characterictics of transport processes in these substances.
\vspace{2mm}

{\bf  Acknowledgments:}
We thank G.M. Zimbovsky for help with the manuscript.
This work was supported  by NSF Advance  program SBE-0123654, PR Space Grant NGTS/40091, and NSF-PREM 0353730.

\vspace{2mm}

\end{document}